# SCIENTIFIC REPORTS

**OPEN**  Simple understanding of quantum weak values

Lupei Qin[1], Wei Feng[2] & Xin-Qi Li[1]

In this work we revisit the important and controversial concept of quantum weak values, aiming to provide a simplified understanding to its associated physics and the origin of *anomaly*. Taking the Stern-Gerlach setup as a working system, we base our analysis on an exact treatment in terms of quantum Bayesian approach. We also make particular connection with a very recent work, where the anomaly of the weak values was claimed from the pure statistics in association with "disturbance" and "post-selection", rather than the unique quantum nature. Our analysis resolves the related controversies through a clear and quantitative way.

The concept of weak values (WVs), introduced by Aharonov, Albert and Vaidman (AAV) nearly 30 years ago[1,2], has caused continuous interests and controversies[3–13]. A large number of references can be found, for instance, in the recent review articles[13–15]. The simple reason for causing both interests and controversies might be seen from the unusual form of the AAV WV:

$$A_w = \frac{\langle \psi_f | \hat{A} | \psi_i \rangle}{\langle \psi_f | \psi_i \rangle}, \qquad (1)$$

where $|\psi_i\rangle$ and $|\psi_f\rangle$ are the pre- and post-selected (PPS) states of $\hat{A}$. The most surprising prediction of this formula is that it can drastically exceed the range of eigenvalues of the observable $\hat{A}$, violating thus our common knowledge.

The AAV WV formula itself, i.e., Eq. (1), was obtained based on quantum mechanics. Therefore, unlike emphasized in refs 10,13, it seems not so surprising that there is no analogous formula based on classical principles which can display the *functional* feature of Eq. (1). However, as properly pointed out in refs 10,13, the appearance of *anomalous* WV based on Eq. (1) (exceeding the range of the eigenvalues of $\hat{A}$) is indeed originated from the *quantum interference*[10,13]. In ref. 13, the anomalous WV is also related with *negative* quasiprobabilities, which highlights further the quantum nature of the AAV WV.

Moreover, in ref. 10, the quantum nature of the AAV WV has been elaborated further as follows. Consider two coupled systems (or the degrees of freedom of a single system), say, "$A$"-plus-"$B$" with coupling Hamiltonian $\lambda \hat{A}\hat{B}$. (In the WV studies, "$B$" is utilized as the meter for quantum measurements). The weak value $A_w$ of $\hat{A}$ is defined by the PPS states of the system "$A$", as given by Eq. (1). $A_w$ plays the role of an effective parameter coupled to $\hat{B}$ and results thus in a "pre-existing" *shift* in the wave function of the system "$B$". This understanding has been highlighted in particular as[10]: "The weak value shifts exist if measured or not, so the weak value is not defined by the statistics of measurement outcomes. The statistical analysis (performed after the post-selection) can just reveal the pre-existing weak values." This particular statement was mainly directed to the recent work by Ferrier and Combes[7], in which (and in the later response article[8]) the *anomaly* of the WV was claimed from a reason of pure statistics associated with *disturbance* and *post-selection*, rather than the unique quantum nature.

Actually, the work by Ferrier and Combes[7] is just the latest of a series of works in the past years on classical analogues of the WVs and associated paradoxes[16] followed by appropriate clarifications[17,18]. Discussions on the classical- *versus*-quantum issues were also put forward in different manners, from aspects such as violation of the Leggett-Garg inequality[19], negative quasiprobability[13], contextual values (contextuality)[20–23], and even the nature of time[24].

In this work we present a simple, explicit, and quite straightforward way to understand how the AAV WV appears as or enters the PPS average of the measurement outcomes, and how the *anomaly* is caused. For the whole problem, two points are essential: one is the post-selection conditioned average; the other is the superposition principle of quantum mechanics. We will base our analysis on an exact treatment in terms of quantum Bayesian approach by taking the Stern-Gerlach setup as a working system. The reason of using Bayesian approach is twofold: (i) it enables to easily obtain the *exact* result (for arbitrary measurement strength) which will serve as the unified starting point for the whole analysis in this work; and (ii) it allows to clarify that any classical model under correct treatment cannot result in anomalous WVs. We notice that this second point does not arrive to full consensus in literature. We believe that the present work can, in a transparent and quantitative way, resolve the recent controversies[7–13].

[1]Center for Advanced Quantum Studies and Department of Physics, Beijing Normal University, Beijing 100875, China. [2]Department of Physics, Tianjin University, Tianjin 300072, China. Correspondence and requests for materials should be addressed to X.-Q.L. (email: lixinqi@bnu.edu.cn)

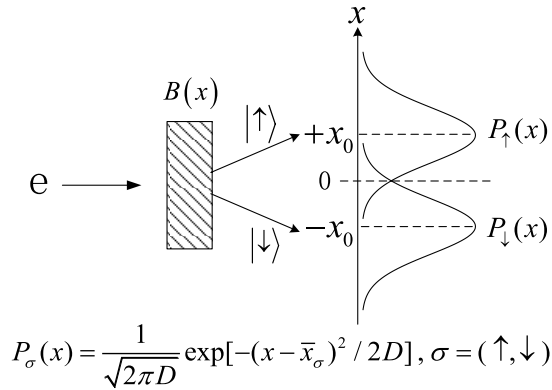

**Figure 1. Schematic plot of the Stern-Gerlach setup utilized for studies of weak measurement and quantum weak values.**

## Results

**AAV's Weak Values.** For the sake of completeness let us briefly review the AAV's treatment of weak values, by taking the Stern-Gerlach setup as a specific working system, as schematically shown in Fig. 1. In this setup the electron's trajectory is deflected when it passing through inhomogeneous magnetic field. Corresponding to "$\lambda \hat{A}\hat{B}$", now the interaction Hamiltonian between the "system" and the "meter" reads $H' = \lambda \hat{A}\hat{p}$. That is, the spin degree of freedom of the electron is the system and the spatial ones (momentum and coordinate) are the meter. In this work we use $\hat{A}$ for the spin operator. Let us consider the system and meter starting the evolution with $|\psi\rangle|\Phi\rangle$ where $|\psi\rangle$ and $|\Phi\rangle$ are, respectively, the system and meter states. For the meter state, i.e., the transverse wavefunction (wavepacket) of the electron, we assume a Gaussian form $\Phi(x) = (2\pi D)^{-1/4} \exp[-x^2/(4D)]$, with $\sqrt{D}$ the width of the wavepacket. In weak coupling limit, which is properly characterized by $\lambda t_m \equiv x_0 \ll \sqrt{D}$ where $t_m$ is the interaction time, short algebra yields[1]

$$\left\langle \phi \middle| \left\langle x \middle| e^{-ix_0\hat{p}\hat{A}} \middle| \psi \right\rangle \middle| \Phi \right\rangle \simeq \langle \phi | \psi \rangle \Phi(x - x_0 A_w), \tag{2}$$

where $A_w = \frac{\langle \phi | \hat{A} | \psi \rangle}{\langle \phi | \psi \rangle}$ is the AAV WV. We see that, indeed, the AAV WV manifests itself as a shift of the wavefunction. It seems that it is largely because of this feature that in ref. 10 the weak value shifts are emphasized as pre-existing. Or, the AAV WV is an effective coupling parameter to the meter system, e.g., via $\lambda A_w \hat{p}$[10]. However, noting the ensemble-statistical interpretation of the quantum wavefunction, we find that this statement is not so different from the opinion by regarding the WV as the statistical average of measurement outcomes. Actually, if we measure the wavefunction in coordinate representation, the measurement outcomes satisfy the statistics with probability $|\Phi(x - x_0 A_w)|^2 \sim \exp[-(x - x_0 \operatorname{Re} A_w)^2/4D]$. In the following, we will see that $\operatorname{Re} A_w$ is the lowest-order approximation of the conditional average of the measurement outcomes associated with the PPS statistics.

**Bayesian Treatment.** To generalize the above analysis from weak coupling limit to finite strength interaction, the best way might be using the quantum Bayesian approach to calculate the PPS conditional average. In ref. 19 this kind of calculation was performed for a solid state qubit measured by quantum-point-contact[25]. Similarly, applying the quantum Bayesian rule for circuit-QED architecture[26,27], the general expression of the associated weak values has been obtained[28].

For the Stern-Gerlach setup, the transverse spatial coordinate of the electron plays the role of a meter which is further collapsed (measured) by an outside classical detector. The probability distribution of the measurement outcomes (the collapsed positions on the screen) is simply given by

$$P_\sigma(x) = |\Phi(x - \bar{x}_\sigma)|^2 = \frac{1}{\sqrt{2\pi D}} \exp\left[-\frac{(x - \bar{x}_\sigma)^2}{2D}\right], \tag{3}$$

where $\sigma = (\uparrow, \downarrow)$, and $\overline{x}_{\uparrow(\downarrow)} = +(-)x_0$ are the distribution centers associated with the states of spin-up $|\uparrow\rangle$ and spin-down $|\downarrow\rangle$.

If $P_\uparrow(x)$ and $P_\downarrow(x)$ are strongly overlapped, the measurement (with outcomes of "$x$") falls into the category of quantum *weak* measurement. In this case, the quantum Bayesian approach is also a perfect tool for the Stern-Gerlach setup. Originally proposed by Korotkov[25], the quantum Bayesian approach is largely based on the well-known Bayes formula in Probability Theory together with a *quantum purity* consideration. The former is utilized to determine the diagonal elements while the latter is for determination of the off-diagonal ones. In quite compact form, one can use the quantum Bayesian rule to update the spin state from $\rho$ to $\tilde{\rho}$ as follows:

$$\begin{aligned}\tilde{\rho}_{\uparrow\uparrow} &= \rho_{\uparrow\uparrow}P_\uparrow(x)/\mathcal{N}(x),\\ \tilde{\rho}_{\downarrow\downarrow} &= \rho_{\downarrow\downarrow}P_\downarrow(x)/\mathcal{N}(x),\\ \tilde{\rho}_{\uparrow\downarrow} &= \rho_{\uparrow\downarrow}\sqrt{P_\uparrow(x)P_\downarrow(x)}/\mathcal{N}(x),\end{aligned} \quad (4)$$

where $\mathcal{N}(x) = \rho_{\uparrow\uparrow}P_\uparrow(x) + \rho_{\downarrow\downarrow}P_\downarrow(x)$ is a normalization factor. Note that the last equality (for the off-diagonal element) stems from the purity consideration[25].

Precisely in parallel to the AAV's treatment, let us consider the PPS states $|\psi\rangle$ and $|\phi\rangle$, or in terms of the density matrices $\rho = |\psi\rangle\langle\psi|$ and $\rho_\phi = |\phi\rangle\langle\phi|$. We will explicitly employ the PPS average of the measurement outcomes as the practical definition of weak values, which is actually in the same spirit of achieving the AAV WV and reads[19–21,28]

$$_\phi\langle x\rangle_\psi = \frac{\int dx\, x P_\psi(x) P_x(\phi)}{\int dx\, P_\psi(x) P_x(\phi)} = \frac{M_1}{M_2}, \quad (5)$$

where $P_\psi(x)$ is the distribution probability of the measurement outcomes with the pre-selected state $\psi$ *before the post-selection*. Note that, actually, $P_\psi(x) = \mathcal{N}(x)$. $P_x(\phi)$ is the post-selection probability which can be obtained via $P_x(\phi) = \text{Tr}[\rho_\phi \tilde{\rho}(x)]$, by applying the quantum Bayesian rule Eq. (4), to update state from $\rho$ to $\tilde{\rho}(x)$ based on the outcome $x$. Obviously, $P_\psi(x)P_x(\phi)$ plays the role of the joint PPS probability of getting "$x$", while having the denominator $M_2$ as its normalization factor. Straightforwardly, by completing a couple of Gaussian integrals, the weak value defined by Eq. (5) is given by

$$\begin{aligned}M_1 &= \left(\rho_{\phi\uparrow\uparrow}\rho_{\uparrow\uparrow}\overline{x}_\uparrow + \rho_{\phi\downarrow\downarrow}\rho_{\downarrow\downarrow}\overline{x}_\downarrow\right) + (\overline{x}_\uparrow + \overline{x}_\downarrow)\\ &\times \text{Re}\left(\rho^*_{\phi\uparrow\downarrow}\rho_{\uparrow\downarrow}\right)\exp[-(\overline{x}_\uparrow - \overline{x}_\downarrow)^2/(8D)],\end{aligned} \quad (6a)$$

$$\begin{aligned}M_2 &= \left(\rho_{\phi\uparrow\uparrow}\rho_{\uparrow\uparrow} + \rho_{\phi\downarrow\downarrow}\rho_{\downarrow\downarrow}\right) + 2\,\text{Re}\left(\rho^*_{\phi\uparrow\downarrow}\rho_{\uparrow\downarrow}\right)\\ &\times \exp[-(\overline{x}_\uparrow - \overline{x}_\downarrow)^2/(8D)].\end{aligned} \quad (6b)$$

Where $\rho_{\sigma\sigma'}$ and $\rho_{\phi\sigma\sigma'}$ are the elements of the density matrices $\rho$ and $\rho_\phi$, respectively.

To establish an explicit connection of the above generalized result with the AAV WV, we need to reexpress the result of Eq. (6). Without loss of generality, let us assume $\overline{x}_{\uparrow,\downarrow} = \pm x_0$. By expressing the AAV WV $A_w = \langle\phi|\hat{A}|\psi\rangle/\langle\phi|\psi\rangle$ in terms of the density matrix elements of $\rho$ and $\rho_\phi$, after some simple algebra we obtain

$$\frac{_\phi\langle x\rangle_\psi}{x_0} = \frac{\text{Re}(A_w)}{1 + \mathcal{G}(|A_w|^2 - 1)}. \quad (7)$$

In this elegant result, we have introduced $\mathcal{G} = (1 - e^{-2g})/2$ and $g = (\overline{x}_\uparrow - \overline{x}_\downarrow)^2/(16D)$.

We see that in the weak measurement limit (small $g$) and with modest $|A_w|$ (not "strange" enough), Eq. (7) returns to the AAV's result. This shows that the AAV's WV (more precisely the real part of it) is indeed the PPS average of the measurement outcomes. Another important feature in the result of Eq. (7) is the second (correction) term in the denominator. It will make the PPS average considerably deviate from the AAV WV for finite strength measurement or with very "strange" $|A_w|$. This feature should be kept in mind when one attempts to extract the AAV WV from the PPS average. Similar result as generalization of the AAV WV has been found as well for qubit measurements by quantum point contact[19] and in the circuit-QED system[28], and has been connected with the more general formulation of contextual values[20,21].

In general, the AAV WV is a complex number. While Eq. (7) relates the PPS average with the real part of the AAV WV, how to relate it with the imaginary part of the AAV WV is of interest. In the context of optical (laser-beam) setup, it was shown in ref. 15 that one can "cleverly" post-select a specific transverse state corresponding to a specific position/momentum of the laser beam, in order to measure the real/imaginary part of the polarization WV. In ref. 29 the meaning and significance of the imaginary part of the AAV WV has been further exploited.

For the Stern-Gerlach setup, if one is able to introduce the "$\hat{x}\hat{A}$"-type interaction in the *system-meter* coupling Hamiltonian, i.e., $H' = \lambda_1\hat{p}\hat{A} + \lambda_2\hat{x}\hat{A}$, then the imaginary part of $A_w$ can appear as well in the numerator of Eq. (7). To be specific, let us assume that the *system-meter*-coupling is switched on for a time interval $t_m$. Then, $x_0$ in the above $\bar{x}_{\uparrow,\downarrow} = \pm x_0$ is given by $x_0 = \lambda_1 t_m \equiv \epsilon_1$. Moreover, when applying the quantum Bayesian rule Eq. (4), a phase factor $e^{-i\epsilon_2 x}$ should be attached to $\tilde{\rho}_{12}$ in the third equality, where $\epsilon_2 = \lambda_2 t_m$. Inserting these accounts into the WV calculations, we obtain

$$_\phi\langle x\rangle_\psi = \frac{\epsilon_1 \text{Re}(A_w) + \epsilon_2 \text{Im}(A_w)}{1 + \mathcal{G}(|A_w|^2 - 1)}, \quad (8)$$

However, for the Stern-Gerlach setup, it seems unclear how to realize the above *dual* coupling Hamiltonian. Alternately, for the circuit-QED system as analyzed in ref. 28, it is indeed possible to obtain the WV expression as Eq. (8). There, even better, one can make either $\epsilon_1 = 0$ or $\epsilon_2 = 0$ by tuning the local oscillator's phase in the homodyne measurement of the cavity field.

**Origin of Anomalies.** Following the standard and practical way of experimentally measuring the quantum average of a physical observable ($\hat{A}$), the above analysis established a general connection (for finite strength measurement) between the PPS average and the quantum AAV WV. From Eq. (7), we see that the extent of anomaly of the PPS average largely depends on $A_w$. Actually, in the weak measurement limit, if we neglect the back-action ("disturbance") effect of the measurement on the measured state, the rescaled PPS average $_\phi\langle x\rangle_\psi/x_0$ is precisely the AAV WV.

As a preliminary and most straightforward illustration, let us first inspect the origin of anomaly of the AAV WV which may be rewritten as

$$\begin{aligned} A_w &= \frac{\langle\phi|\hat{A}|\psi\rangle}{\langle\phi|\psi\rangle} = \frac{\langle\phi|\hat{A}|\psi\rangle\langle\psi|\phi\rangle}{\langle\phi|\psi\rangle\langle\psi|\phi\rangle} \\ &= \frac{\text{Tr}[\hat{A}\rho\rho_\phi]}{\text{Tr}[\rho\rho_\phi]} = \frac{M_1}{M_2} \end{aligned} \quad (9)$$

To be specific, consider the measurement of the real part of $A_w$. We have $\text{Re}(M_1) = \rho_{\uparrow\uparrow}\rho_{\phi\uparrow\uparrow} - \rho_{\downarrow\downarrow}\rho_{\phi\downarrow\downarrow}$ and $M_2 = |\langle\phi|\psi\rangle|^2$. Let us also specify the PPS states as

$$\begin{aligned} \psi &= \alpha|\uparrow\rangle + \beta|\downarrow\rangle, \\ \phi &= a|\uparrow\rangle + b|\downarrow\rangle, \end{aligned} \quad (10)$$

and assume that all the superposition coefficients ($\alpha$, $\beta$) and ($a$, $b$) are real. Then we obtain $M_1 = a^2\alpha^2 - b^2\beta^2$ and $M_2 = (a\alpha + b\beta)^2$, and achieve the divergence condition $a\alpha + b\beta \to 0$, which corresponds to an ultra-small post-selection probability. In this case we also have $M_1 \to 0$ which, however, is a first-order small quantity while $M_2$ is of the second order. This is essentially equivalent to the divergence feature of the AAV weak value $A_w$, i.e., $\langle\phi|\hat{A}|\psi\rangle \neq 0$ when $\langle\phi|\psi\rangle \to 0$.

Then, we see that it is right the *quantum interference* that possibly makes $|a\alpha + b\beta|^2 \to 0$, while $|a|^2|\alpha|^2 + |b|^2|\beta|^2 \neq 0$. In classical case, the superposed *amplitudes* in the quantum states $|\psi\rangle$ and $|\phi\rangle$ should be replaced by *probabilities* (modulus squares of the amplitudes). This would result in the joint PPS probability given by $M_2 = |a|^2|\alpha|^2 + |b|^2|\beta|^2$. Therefore, we can definitely conclude that in classical case, the PPS average of $\hat{A}$ is *impossible* to exceed the normal bounds of $[-1, 1]$, i.e., there is no *anomalous* classical weak values.

Now consider the finite strength measurement. Let us rewrite the joint PPS probability $M_2$ in Eq. (6) as $M_2 = |\langle\phi|\psi\rangle|^2 + \delta M_2$, where

$$\delta M_2 = 2 \text{Re}\left(\rho^*_{\phi\uparrow\downarrow}\rho_{\uparrow\downarrow}\right)[e^{-(\bar{x}_\uparrow - \bar{x}_\downarrow)^2/8D} - 1]. \quad (11)$$

This "correction" (to the probability $|\langle\phi|\psi\rangle|^2$) is originated from the "disturbance" (back-action) of the quantum measurement on the pre-selected state $|\psi\rangle$, which alters the PPS probability by the amount of $\delta M_2$. Notably, in a sharp contrast to the classical model to be discussed in the following, this "disturbance" is *not* at all the origin of causing anomalous weak values. Actually, owing to the presence of $\delta M_2$, the PPS probability cannot approach zero. This implies that the WV is not to be divergent when $|\langle\phi|\psi\rangle|^2 \to 0$. In contrast, noting that $\text{Re}(M_1) \to 0$ in this case, the presence of the nonzero $\delta M_2$ will result in a vanishing WV[28]. The reason is that the "disturbance" can result in successful post-selection while it is impossible (owing to destructive quantum interference) if there is no disturbance. Finally we point out that, for finite strength measurement (in the presence of $\delta M_2$, there still exists a window of post-selection and measurement strength for the appearance of *anomalous* WVs.

**Classical Coin-Toss Model.** In the recent articles by Ferrier and Combes[7,8], it was claimed that the anomalous weak values are not uniquely relate to quantum nature, but rather a purely statistical feature of pre- and post-selection with "disturbance". Below we briefly revisit the classical *coin-toss* model analyzed in refs 7,8, in attempt to provide a simple view for the origin of the "anomaly" displayed there.

The coin-toss model, which was originally analyzed in ref. 7, is actually a *coarse-grained* version of the Stern-Gerlach setup which we have discussed above. The probability of the coarse-grained outcome distribution of the "weak" measurements was proposed as[7]

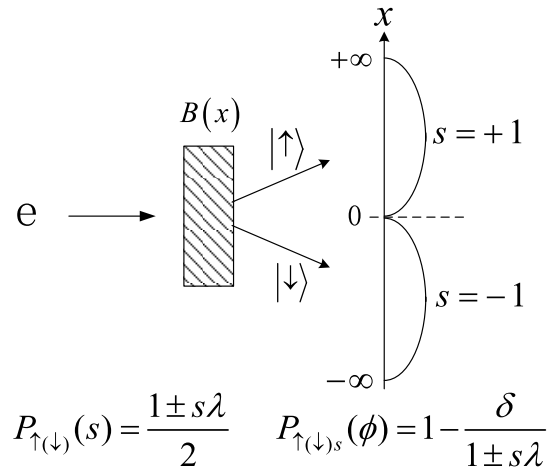

$$P_{\uparrow(\downarrow)}(s) = \frac{1 \pm s\lambda}{2} \qquad P_{\uparrow(\downarrow)s}(\phi) = 1 - \frac{\delta}{1 \pm s\lambda}$$

**Figure 2. The coarse-grained version of the Stern-Gerlach setup which corresponds to the coin-toss model discussed in ref. 7, by regarding all the outcomes** $x \in (0, \pm \infty)$ **as** $s = \pm 1$. $P_{\uparrow(\downarrow)}(s)$ is the "coarse-grained" spatial distribution probability of the "spin-up(down)" state, while $P_{\uparrow(\downarrow)s}(\phi)$ is its subsequent post-selection probability (note that it has an unusual/artificial feature of being $s$-dependent). $\phi$ is the post-selection state which was chosen as $\phi = |\downarrow\rangle$ in ref. 7 (and in the present work), but in general which can be a classical mixture of $|\uparrow\rangle$ and $|\downarrow\rangle$.

$$P_\psi(s) = \frac{1}{2}(1 + s\lambda \bar{A}_\psi), \tag{12}$$

where $\bar{A}_\psi = \langle\psi|A|\psi\rangle$ and $s = \pm 1$. Here, the coarse-grained variable "$s = \pm 1$" correspond to the integrated outcomes of $x \in (0, \pm\infty)$. Note that in ref. 7, this coarse-grained variable was used in certain confusing manner together with the "Heads" and "Tails" in the coin-toss model, which hides then the serious artificial feature of post-selection rule assumed there. Below, to avoid such type of ambiguity, we use the terms of "spin-up" and "spin-down" for the "Heads" and "Tails", i.e., the *intrinsic* coin states, while emphasizing $s = \pm 1$ for the meter's coarse-grained outputs, as particularly shown by the plot of Fig. 2.

Now, more specifically, for the "spin-up" and "spin-down" coins, the coarse-grained locations obey the integrated probabilities in the two regions given by, respectively,

$$P_\uparrow(s) = (1 + s\lambda)/2, \tag{13a}$$

$$P_\downarrow(s) = (1 - s\lambda)/2. \tag{13b}$$

From these, one may clearly keep in mind that the "spin-up" and "spin-down" coins would locate in the "$s = +1$" and "$s = -1$" regions with different probabilities. One can easily check $\bar{s} = \sum_{s=\pm 1} s P_\psi(s) = \lambda \bar{A}_\psi$. This implies that $\lambda$ is the scaling parameter between the quantum expectation of $\hat{A}$ and the data average of meter's outcomes. One can thus reasonably regard it as the measurement strength.

The key step in the WV analysis of the coin-toss model is to introduce a "disturbance" (bit-flip channel), which is modeled by

$$|\uparrow\rangle\langle\uparrow| \mapsto (1 - P_{\uparrow s})|\uparrow\rangle\langle\uparrow| + P_{\uparrow s}|\downarrow\rangle\langle\downarrow| \tag{14a}$$

$$|\downarrow\rangle\langle\downarrow| \mapsto (1 - P_{\downarrow s})|\uparrow\rangle\langle\uparrow| + P_{\downarrow s}|\downarrow\rangle\langle\downarrow| \tag{14b}$$

where

$$P_{\uparrow(\downarrow)s} = 1 - \frac{\delta}{1 \pm s\lambda} \equiv P_{\uparrow(\downarrow)s}(\phi). \tag{15}$$

Note that in ref. 7 only the first process was explicitly displayed, because of $\psi = |\uparrow\rangle$ considered there. Obviously, $\delta$ characterizes the amount of disturbance and $P_{\uparrow(\downarrow)s}$ corresponds to the post-selection probability of $\phi = |\downarrow\rangle$ (hereafter we assume the post-selection state of $\phi = |\downarrow\rangle$, as the same in ref. 7).

To uncover the underlying problem more transparently, we would like to present a slightly generalized treatment by considering a superposition pre-selected state $|\psi\rangle = c_\uparrow|\uparrow\rangle + c_\downarrow|\downarrow\rangle$, instead of $|\psi\rangle = |\uparrow\rangle$ as in[7]. Now, applying the PPS average scheme

www.nature.com/scientificreports/

$$_\phi \langle s \rangle_\psi = \frac{\sum_{s=\pm 1} \sum_{\sigma=\uparrow,\downarrow} s P_{\psi,\sigma}(s) P_{\sigma s}(\phi)}{\sum_{s=\pm 1} \sum_{\sigma=\uparrow,\downarrow} P_{\psi,\sigma}(s) P_{\sigma s}(\phi)} = \frac{M_1}{M_2}, \quad (16)$$

where $P_{\psi,\sigma}(s) = |c_\sigma|^2 P_\sigma(s)$, after simple algebra we obtain the same result of weak value as derived in ref. 7:

$$A_w = \frac{_\phi \langle s \rangle_\psi}{\lambda} = \frac{\bar{A}_\psi}{1-\delta}. \quad (17)$$

Remarkably, this classical WV can become "anomalous" and even be very "strange" by altering the parameter $\delta$. This is the key result of ref. 7.

**Comparative Analysis.** Based on the Bayesian treatment, we can convert the quantum result to its classical counterpart by simply dropping the off-diagonal terms in $M_1$ and $M_2$ in Eq. (6):

$$M_1 = \int dx\, x\Big[\rho_{\phi\uparrow\uparrow} P_\uparrow(x) \rho_{\uparrow\uparrow} + \rho_{\phi\downarrow\downarrow} P_\downarrow(x) \rho_{\downarrow\downarrow}\Big], \quad (18a)$$

$$M_2 = \int dx\Big[\rho_{\phi\uparrow\uparrow} P_\uparrow(x) \rho_{\uparrow\uparrow} + \rho_{\phi\downarrow\downarrow} P_\downarrow(x) \rho_{\downarrow\downarrow}\Big]. \quad (18b)$$

Accordingly, we obtain $M_1 = (\rho_{\phi\uparrow\uparrow} \rho_{\uparrow\uparrow} - \rho_{\phi\downarrow\downarrow} \rho_{\downarrow\downarrow})\lambda$ and $M_2 = \rho_{\phi\uparrow\uparrow} \rho_{\uparrow\uparrow} + \rho_{\phi\downarrow\downarrow} \rho_{\downarrow\downarrow}$. We see then that in classical case the PPS average cannot become *anomalous* since we always have $|M_1/M_2| \leq \lambda$.

Now let us reformulate the coin-toss model via

$$M_1 = \sum_{s=\pm 1} s\Big[\rho_{\phi\uparrow\uparrow s} P_\uparrow(s) \rho_{\uparrow\uparrow} + \rho_{\phi\downarrow\downarrow s} P_\downarrow(s) \rho_{\downarrow\downarrow}\Big], \quad (19a)$$

$$M_2 = \sum_{s=\pm 1} \Big[\rho_{\phi\uparrow\uparrow s} P_\uparrow(s) \rho_{\uparrow\uparrow} + \rho_{\phi\downarrow\downarrow s} P_\downarrow(s) \rho_{\downarrow\downarrow}\Big]. \quad (19b)$$

In the classical coin-toss analysis[7], the most problematic procedure is the "insertion" of the following post-selection rule (the so-called "noisy channel" or "disturbance"):

$$\rho_{\phi\uparrow\uparrow s} = 1 - \frac{\delta}{1+s\lambda}, \quad \rho_{\phi\downarrow\downarrow s} = 1 - \frac{\delta}{1-s\lambda}. \quad (20)$$

Accordingly, one gets $M_1 = \lambda \rho_{\uparrow\uparrow} + (-\lambda)\rho_{\downarrow\downarrow} = \lambda \bar{A}_\psi$ and $M_2 = (1-\delta)(\rho_{\uparrow\uparrow} + \rho_{\downarrow\downarrow}) = 1 - \delta$. Then, by means of this procedure, one obtains anomalous WV since the scaled PPS average "$\frac{M_1}{M_2}/\lambda$" can drastically exceed $\bar{A}_\psi$.

Some remarks on the above post-selection "rule" are in order as follows. (i) Indeed, the overall post-selection probability $\langle\phi|\tilde{\rho}(x)|\phi\rangle$ depends on the outcome "$x$" (or the coarse-grained location "$s$"). However, in either quantum or classical weak ("noisy") measurement, this dependence has been fully accounted for by the Bayesian rule, via updating the state from $\rho$ to $\tilde{\rho}(x)$. Then the post-selection probabilities from the component (basis) states $|\uparrow\rangle$ and $|\downarrow\rangle$, say, $|\langle\phi|\uparrow\rangle|^2$ and $|\langle\phi|\downarrow\rangle|^2$, should no longer depend on "$x$" or "$s$". In contrast, the post-selection probabilities given by Eq. (20) depend on $s=+1$ or $s=-1$. This is a misleading procedure to "generate" the anomalous classical WV. The "rule" of Eq. (20), or any other "$s$-dependent rule", is an artifact, which simply means keeping or discarding the stochastic events ($s=\pm 1$) according to our willing, then re-calculating the average of $s$ and extracting the weak values as $_\phi\langle s\rangle_\psi/\lambda$. Obviously, it does not make sense by comparing this type of PPS average with the quantum WV. (ii) So we should obey the *convention* that the post-selection probabilities $|\langle\phi|\uparrow\rangle|^2$ and $|\langle\phi|\downarrow\rangle|^2$ do *not* depend on "$x$" or "$s$". Under this requirement, as proved below Eq. (18), we conclude that it is *impossible* to generate anomalous WV in any classical contexts. (iii) In order to get anomalous WV, the only way is adding the interference terms into $M_2$ [c.f. Eqs. (6) and (18)], to make $|M_1/M_2| < \lambda$. This is possible only in quantum case. So the anomalous WV is indeed a *unique* quantum phenomenon, which deeply originates from quantum interference, or the most fundamental quantum superposition principle. Actually, it was proved in ref. 19 that the anomalous WV is equivalent to the violation of the Leggett-Garg inequality, which is also a direct consequence of the quantum superposition principle. (iv) The anomalous WV is owing to the distortion of the joint PPS probability distribution, which is caused by quantum interference in quantum system but in the classical coin-toss model by an artificial procedure. In ref. 7 the "post-selection" was termed as "disturbance" (or "noisy channel"). However, for the problem under consideration, the only acceptable "disturbance" is the measurement (or information-gain) backaction, which has been fully accounted for by the Bayesian rule, for both the quantum and classical measurements. As clearly seen in the quantum WV analysis in this work, this type of "disturbance" would reduce the "anomaly" amount, which is in sharp contrast with the coin-toss model where the extra "disturbance" is the key reason of generating the "anomaly".

## Discussion

To summarize, we have presented a simple and direct method to revisit the concept of quantum weak values. The Bayesian treatment associated analysis in comparison with a controversial classical model supports the assertion that the anomalous weak values are purely quantum mechanical, having no classical analogue. That is, in addition to the argument of *functional* dependence[10,13], we arrive to a stronger conclusion: the anomalous WVs cannot be reproduced by any *correctly* treated classical model.

This conclusion is in full agreement with the finding uncovered in ref. 19, where the equivalence proof between the anomalous WV and the violation of Leggett-Garg inequality implies that the anomalous WVs rule out any classical (hidden-variable) interpretation. Therefore, an insistence that disallows adding extra disturbance is to make the classical-quantum comparison at equal foot. The reason is just like the following: if one introduces "extra procedures" in the (classical) hidden-variable treatment, it would destroy the meaning of violation of the Bell-Leggett-Garg inequalities.

In quantum mechanics, the wave function is actually a "knowledge" which is to be altered after measurement. This is the so-called backaction or disturbance of quantum measurement. For the *noisy* measurement in the classical coin-toss model, the measurement outcome will also change the prior probability ("knowledge") known before the measurement. So in this sense a noisy classical measurement resembles the quantum measurement — both obey the Bayesian rule. This explains further that in any classical model, a correct treatment should disallow adding extra "disturbance", since the classical "information-gain backaction", which corresponds to the quantum measurement backaction, has been accounted for by the Bayesian rule, as clearly analyzed in our work by Eqs. (18) and (19).

## Acknowledgements

This work was supported by the NNSF of China under grant No. 10874176 and the State "973" Project under grant No. 2012CB932704.

## Author Contributions

X.-Q.L. supervised the work. L.Q. and W.F. carried out the calculations. X.-Q.L. wrote the paper and all authors reviewed it.

### Additional Information

**Competing financial interests:** The authors declare no competing financial interests.

**How to cite this article**: Qin, L. *et al.* Simple understanding of quantum weak values. *Sci. Rep.* **6**, 20286; doi: 10.1038/srep20286 (2016).